\documentclass[aps,prx,twocolumn,groupedaddress,amsmath,epsfig,amssymb,eqsecnum]{revtex4-1}
\usepackage{mathrsfs}
\usepackage{makecell}
\usepackage{bbm}
\usepackage{bm}
\usepackage{amsfonts}
\usepackage{amsthm}
\usepackage{color}
\usepackage{graphicx}
\usepackage{booktabs}
\usepackage{multirow}
\usepackage{tikz}
\usetikzlibrary{arrows.meta, hobby}
\usepackage{enumitem}
\usepackage{braket}
\newtheorem{definition}{Definition}
\newtheorem{proposition}{Proposition}

\newcommand{\be}{\begin{equation}}
\newcommand{\ee}{\end{equation}}
\newcommand{\ba}{\begin{eqnarray}}
\newcommand{\ea}{\end{eqnarray}}

\newcommand{\dif}{\mathrm{d}}
\newcommand{\B}{\mathcal{B}}

\newcommand{\Fboson}{\mathcal{F}_{\text{boson}}}
\newcommand{\Hqubit}{\mathcal{H}_{\text{qubit}}}
\usepackage{hyperref}
\hypersetup{
colorlinks=true,
linkcolor=blue,
urlcolor=blue,
citecolor=blue,
}

\begin{document}
\title{Geometry of Quantum Logic Gates}
\author{M.W. AlMasri}
\email{mwalmasri2003@gmail.com}
\affiliation{Wilczek Quantum Center, School of Physics and Astronomy, Shanghai Jiaotong University, Minhang, Shanghai, China}
\begin{abstract}
In this work, we investigate the geometry of quantum logic gates within the holomorphic representation of quantum mechanics. We begin by embedding the physical qubit subspace into the space of holomorphic functions that are homogeneous of degree one in each Schwinger boson pair $(z_{a_j}, z_{b_j})$. Within this framework, we derive explicit closed-form differential operator representations for a universal set of quantum gates, including the Pauli operators, Hadamard, CNOT, CZ, and SWAP, and demonstrate that they preserve the physical subspace exactly. Restricting to unit-magnitude variables ($|z| = 1$) reveals a toroidal space $\mathbb{T}^{2N}$, on which quantum gates act as canonical transformations: Pauli operators generate Hamiltonian flows, the Hadamard gate induces a nonlinear automorphism, and entangling gates produce correlated diffeomorphisms that couple distinct toroidal factors. Beyond the torus, the full Segal--Bargmann space carries a natural Kähler geometry that governs amplitude dynamics. Entanglement is geometrically characterized via the Segre embedding into complex projective space, while topological protection arises from the $U(1)^N$ fiber bundle structure associated with the Jordan--Schwinger constraint.
\end{abstract}
\maketitle

\section{Introduction}
The representation of quantum information in continuous phase space has long provided profound insights into the structure of quantum theory, from the Wigner--Weyl correspondence to modern formulations of quantum optics and quantum computation \cite{Strocchi,Schleich,Zachos,review}. In \cite{phase}, the authors represented both the states and the evolution of a quantum computer in phase space using the discrete Wigner function. This framework allows us to analyze quantum algorithms, such as the Fourier transform and Grover's search, and to examine the conditions under which quantum evolution corresponds directly to classical dynamics in phase space. The Bargmann representation provides a holomorphic (complex-analytic) realization of quantum states in which wavefunctions are entire functions of a complex variable $z \in \mathbb{C}$, related to the phase-space coordinates via
$
z = \frac{q + i p}{\sqrt{2}}
$
(in dimensionless units). In this representation, the standard phase space of classical mechanics, spanned by position $q$ and momentum $p$, is identified with the complex plane, and quantum states correspond to square-integrable holomorphic functions with respect to the Gaussian measure $e^{-|z|^2} \, d^2z$. Coherent states map to simple monomials or exponentials, and operators act as differential operators, making the Segal--Bargmann space particularly well-suited for studying bosonic systems, semiclassical limits, and analytic properties of quantum dynamics. Thus, the Bargmann representation offers a powerful bridge between Hilbert space quantum mechanics and complex phase space geometry \cite{Segal,Bargmann,Perelomov, Folland,Bishop, Hall,almasri,almasri1}.
The space of quantum rays naturally carries a \textit{K\"ahler} manifold structure, enabling a geometric reformulation of quantum mechanics in which states correspond to points on a symplectic manifold equipped with a compatible Riemannian metric~\cite{KahlerQC}. Within this framework, observables are realized as real-valued functions and Schr\"odinger evolution emerges as a Hamiltonian symplectic flow, closely paralleling classical mechanics. Intrinsically quantum phenomena, such as uncertainty relations and state reduction, instead originate from the Riemannian component, which lacks a classical analogue. This perspective clarifies foundational issues including second quantization, coherent-state semiclassics, and the WKB approximation, while establishing a unified setting for dynamical generalizations of quantum theory.

Complementing this general foundation, \cite{ComplexGeoQC} applies the Fubini--Study geometry to low-spin Hilbert spaces, mapping physical observables and symmetries to intrinsic geometric features. This approach yields a canonical measure of entanglement, identifies Schr\"odinger trajectories with Killing vector fields and their associated geometric phases, and quantifies measurement uncertainty through the gradient structure of expectation-value surfaces, enabling systematic corrections to the Heisenberg principle. By further characterizing mixed states as probability distributions whose first moments reconstruct density operators, the work establishes a comprehensive geometric framework that naturally extends standard quantum formalism.

Extending these geometric and analytic insights to bosonic systems, Chabaud et al.~\cite{Chabaud} employ the Segal--Bargmann holomorphic representation to bridge discrete- and continuous-variable quantum information. Their central result demonstrates that Gaussian unitary evolution decomposes exactly into an integrable Calogero--Moser dynamical system governing the motion of zeros of the holomorphic wavefunction, coupled to a conformal flow of Gaussian covariance parameters. Leveraging this correspondence, they construct a systematic non-Gaussian state hierarchy, relate multimode entanglement to the factorization properties of entire functions, and derive a unified classification of subuniversal quantum computational models that generalizes both Boson Sampling and Gaussian quantum computing.
\vskip 5mm
In this work, we construct an explicit and self-contained holomorphic representation of quantum logic gates by unifying two fundamental mappings: (i) the Schwinger boson representation, which encodes each spin into a pair of bosonic modes subject to a total-occupation constraint; and (ii) the Segal--Bargmann transform, which maps the bosonic Fock space to a space of holomorphic functions. We show that any $N$-qubit gate can be expressed as a differential operator acting on Segal--Bargmann space functions that are homogeneous of degree one in each Schwinger boson pair $(z_{a_j}, z_{b_j})$. This representation preserves the physical subspace exactly and provides a direct link between algebraic gate operations and geometric flows on phase space.

Our key contributions are threefold:
(i) We derive closed-form differential operator expressions for fundamental single- and multi-qubit gates, including Pauli operators, Hadamard, SWAP, CNOT, and CZ, in terms of holomorphic variables.
(ii) We analyze the induced dynamics on the toroidal space $\mathbb{T}^{2N}$ obtained by restricting to unit-magnitude variables ($|z| = 1$), revealing that quantum gates act as canonical transformations with distinct geometric signatures (e.g., Hamiltonian flows for Pauli gates, nonlinear automorphisms for Hadamard).
(iii) We uncover deeper geometric structures beyond the torus: the full Segal--Bargmann space carries a Kähler geometry governing amplitude dynamics; entanglement is characterized via the Segre embedding; and topological protection emerges from the $U(1)^N$ fiber bundle structure associated with the Jordan--Schwinger constraint.

The remainder of this paper is structured as follows. In Sec.~\ref{sec:phase_space}, we present the foundational mapping from fermionic systems to holomorphic representation. We derive the homogeneity constraint that characterizes physical qubit states and show how computational basis states are encoded in holomorphic variables.
Section~\ref{sec:gates} presents the core result: explicit differential operator representations of universal quantum logic gates in the Segal--Bargmann space. We provide closed-form expressions for single-qubit gates (Pauli operators, Hadamard) and multi-qubit gates (SWAP, CNOT, CZ), demonstrating that all preserve the physical subspace.
In Sec.~\ref{sec:phasor}, we restrict to unit-magnitude variables ($|z| = 1$) to reveal the toroidal space $\mathbb{T}^{2N}$, where quantum gates act as canonical transformations. We analyze Pauli gates as Hamiltonian flows, Hadamard as a nonlinear automorphism, and entangling gates as diffeomorphisms that correlate distinct toroidal factors.
Section~\ref{sec:beyond_torus} extends the analysis beyond the torus to uncover deeper geometric structures: the Kähler geometry governing amplitude dynamics, the Segre embedding for entanglement characterization, the fiber bundle structure underlying topological protection, and the path integral formulation for semiclassical simulation.
Finally, Sec.~\ref{sec:conclusion} summarizes our findings and discusses implications for quantum simulation, error analysis, and geometric quantum computing.

\section{Qubits in Holomorphic Representation}\label{sec:phase_space}
We consider a system of $N$ qubits, where each qubit $j$ is described by the Pauli operators $\sigma_j^x, \sigma_j^y, \sigma_j^z$ satisfying the $\mathfrak{su}(2)$ algebra
\begin{align}
[\sigma_j^\alpha, \sigma_k^\beta] &= 2i\,\delta_{jk}\,\epsilon_{\alpha\beta\gamma}\,\sigma_j^\gamma, \\
\{\sigma_j^\alpha, \sigma_j^\beta\} &= 2\,\delta_{\alpha\beta}\,\mathbb{I},
\end{align}
for $\alpha,\beta,\gamma \in \{x,y,z\}$. The computational basis states $\ket{0}_j$ and $\ket{1}_j$ are eigenstates of $\sigma_j^z$ with eigenvalues $+1$ and $-1$, respectively.

Each qubit is encoded into a pair of bosonic modes $(a_j, b_j)$ via the Schwinger boson representation \cite{Schwinger}:
\begin{align}
\sigma_j^+ &= a_j^\dagger b_j, \label{eq:schwinger_plus} \\
\sigma_j^- &= b_j^\dagger a_j, \label{eq:schwinger_minus} \\
\sigma_j^z &= a_j^\dagger a_j - b_j^\dagger b_j, \label{eq:schwinger_z}
\end{align}
where $\sigma_j^\pm = (\sigma_j^x \pm i\sigma_j^y)/2$. The bosonic operators satisfy canonical commutation relations $[a_j, a_k^\dagger] = [b_j, b_k^\dagger] = \delta_{jk}$, with all other commutators vanishing.

Physical qubit states are restricted to the subspace of total occupation number one for each mode pair:
\begin{equation}
\hat{n}_j^{\mathrm{tot}} \equiv a_j^\dagger a_j + b_j^\dagger b_j = 1.
\label{eq:occupation_constraint}
\end{equation}
This constraint ensures that the Schwinger mapping \eqref{eq:schwinger_plus}--\eqref{eq:schwinger_z} faithfully reproduces the spin-$\tfrac{1}{2}$ algebra. Consequently, the $N$-qubit Hilbert space $\Hqubit = (\mathbb{C}^2)^{\otimes N}$ is isomorphic to the physical subspace $\mathcal{H}_{\mathrm{boson}}^{\mathrm{phys}}$ of the bosonic Fock space $\Fboson$ for $2N$ modes, defined by
\begin{equation}
\hat{n}_j^{\mathrm{tot}} \ket{\psi} = \ket{\psi} \quad \forall\, j = 1, \dots, N.
\end{equation}

The bosonic Fock space $\Fboson$ is mapped to the Segal--Bargmann space $\mathcal{H}_{\text{SB}}$, consisting of holomorphic functions $f : \mathbb{C}^{2N} \to \mathbb{C}$ that are square-integrable with respect to the Gaussian measure
\begin{equation}
\dif\mu(\mathbf{z}) = e^{-\|\mathbf{z}\|^2} \prod_{j=1}^N \frac{\dif^2 z_{a_j} \, \dif^2 z_{b_j}}{\pi^2},
\end{equation}
where $\|\mathbf{z}\|^2 = \sum_{j=1}^N (|z_{a_j}|^2 + |z_{b_j}|^2)$. The Bargmann correspondence is defined by:
\begin{itemize}[left=0pt]
\item Vacuum state: $\ket{0} \mapsto f_0(\mathbf{z}) = 1$,
\item Creation operators: $a_j^\dagger \mapsto z_{a_j}$, \; $b_j^\dagger \mapsto z_{b_j}$,
\item Annihilation operators: $a_j \mapsto \partial / \partial z_{a_j}$, \; $b_j \mapsto \partial / \partial z_{b_j}$.
\end{itemize}

The physical constraint \eqref{eq:occupation_constraint} translates into a homogeneity condition on the holomorphic functions:
\begin{equation}
\left( z_{a_j} \frac{\partial}{\partial z_{a_j}} + z_{b_j} \frac{\partial}{\partial z_{b_j}} \right) f(\mathbf{z}) = f(\mathbf{z}) \quad \forall\, j.
\label{eq:homogeneity}
\end{equation}
Hence, physical qubit states correspond to functions that are homogeneous of degree one in each bosonic pair $(z_{a_j}, z_{b_j})$:
\begin{equation}
f(\mathbf{z}) = \sum_{\mathbf{s} \in \{0,1\}^N} c_{\mathbf{s}} \prod_{j=1}^N z_{a_j}^{\,1 - s_j} z_{b_j}^{\,s_j},
\label{eq:holomorphic_expansion}
\end{equation}
where $s_j \in \{0,1\}$ labels the computational basis state of qubit $j$ ($s_j = 0$ for $\ket{0}_j$, $s_j = 1$ for $\ket{1}_j$), and $c_{\mathbf{s}} \in \mathbb{C}$ are complex expansion coefficients satisfying $\sum_{\mathbf{s}} |c_{\mathbf{s}}|^2 = 1$. The multi-index $\mathbf{s} = (s_1, \dots, s_N)$ runs over all $2^N$ computational basis states.

In this representation, the logical basis states map to simple monomials:
\begin{align}
\ket{0}_j &\mapsto z_{a_j}, \label{eq:zero_mapping} \\
\ket{1}_j &\mapsto z_{b_j}. \label{eq:one_mapping}
\end{align}
Superpositions and entangled states correspond to linear combinations of such monomials, with the homogeneity condition \eqref{eq:homogeneity} ensuring that the total degree in each pair $(z_{a_j}, z_{b_j})$ remains exactly one.

\section{Holomorphic Representation of Quantum Logic Gates}\label{sec:gates}
Following the construction outlined above, we now express fundamental quantum logic gates, namely the Pauli gates $X$, $Y$, $Z$, the Hadamard gate $H$, and the two-qubit SWAP gate, as differential operators acting on holomorphic functions $f(\mathbf{z}) \in \B$ that satisfy the homogeneity constraint
\begin{equation}
\left( z_{a_j} \partial_{z_{a_j}} + z_{b_j} \partial_{z_{b_j}} \right) f = f \quad \forall j.
\end{equation}
In this representation, the logical basis states correspond to:
\begin{align}
\ket{0}_j \equiv \ket{\uparrow}_j &\mapsto z_{a_j}, \\
\ket{1}_j \equiv \ket{\downarrow}_j &\mapsto z_{b_j}.
\end{align}
\subsection{Single-Qubit Gates}
Using the Schwinger boson mapping $\sigma_j^+ = a_j^\dagger b_j$, $\sigma_j^- = b_j^\dagger a_j$, $\sigma_j^z = a_j^\dagger a_j - b_j^\dagger b_j$, and applying the Bargmann correspondence, we obtain:
\begin{itemize}[left=0pt]
\item \textbf{Pauli-$X$ (bit-flip)}:
\begin{equation}
X_j \;\mapsto\; z_{a_j} \partial_{z_{b_j}} + z_{b_j} \partial_{z_{a_j}}.
\end{equation}
\item \textbf{Pauli-$Y$ (bit-phase flip)}:
\begin{equation}
Y_j \;\mapsto\; -i\left( z_{a_j} \partial_{z_{b_j}} - z_{b_j} \partial_{z_{a_j}} \right).
\end{equation}
\item \textbf{Pauli-$Z$ (phase flip)}:
\begin{equation}
Z_j \;\mapsto\; z_{a_j} \partial_{z_{a_j}} - z_{b_j} \partial_{z_{b_j}}.
\end{equation}
\item \textbf{Hadamard gate}:
The Hadamard gate acts as a linear transformation on the holomorphic coordinates:
\begin{equation}
(H_j f)(\dots, z_{a_j}, z_{b_j}, \dots) = f\left(\dots, \tfrac{z_{a_j} + z_{b_j}}{\sqrt{2}}, \tfrac{z_{a_j} - z_{b_j}}{\sqrt{2}}, \dots \right).
\end{equation}
This is equivalent to the pullback under the orthogonal map $(z_{a_j}, z_{b_j}) \mapsto \big( (z_{a_j} + z_{b_j})/\sqrt{2}, (z_{a_j} - z_{b_j})/\sqrt{2} \big)$.
\item \textbf{Two-qubit SWAP gate}:
The SWAP gate exchanges the holomorphic variables of two qubits:
\begin{multline}
(\mathrm{SWAP}_{jk} f)(\dots, z_{a_j}, z_{b_j}, \dots, z_{a_k}, z_{b_k}, \dots) \\
= f(\dots, z_{a_k}, z_{b_k}, \dots, z_{a_j}, z_{b_j}, \dots).
\end{multline}
This permutation preserves the physical homogeneity condition and implements $\mathrm{SWAP}_{jk} \ket{s_j, s_k} = \ket{s_k, s_j}$.
\end{itemize}
\subsection{Multi-Qubit Gates}
Multi-qubit gates are constructed by combining single-qubit operators across distinct mode pairs. Since each qubit $j$ is encoded in its own bosonic pair $(z_{a_j}, z_{b_j})$, gates acting on multiple qubits factorize naturally in the holomorphic representation.
\begin{itemize}[left=0pt]
\item \textbf{Controlled-NOT (CNOT$_{c,t}$)}:
The CNOT gate flips the target qubit $t$ if the control qubit $c$ is in state $\ket{1}$. Using the projector $\ket{1}\bra{1}_c = \frac{1}{2}(1 - Z_c)$, we write:
\begin{equation}
\mathrm{CNOT}_{c,t} = \frac{1 + Z_c}{2} + \frac{1 - Z_c}{2} \, X_t.
\end{equation}
In Segal-Bargmann space $\mathcal{H}_{\text{SB}}$, this becomes:
\begin{multline}
\mathrm{CNOT}_{c,t} \;\mapsto\;
\tfrac{1}{2} \left( 1 + z_{a_c} \partial_{z_{a_c}} - z_{b_c} \partial_{z_{b_c}} \right) \\
+ \tfrac{1}{2} \left( 1 - z_{a_c} \partial_{z_{a_c}} + z_{b_c} \partial_{z_{b_c}} \right)
\left( z_{a_t} \partial_{z_{b_t}} + z_{b_t} \partial_{z_{a_t}} \right).
\end{multline}
\item \textbf{Controlled-Z (CZ$_{c,t}$)}:
The CZ gate applies a phase flip when both qubits are $\ket{1}$:
\begin{equation}
\mathrm{CZ}_{c,t} = \tfrac{1}{2} \left( 1 + Z_c + Z_t - Z_c Z_t \right).
\end{equation}
Its Bargmann representation is:
\begin{multline}
\mathrm{CZ}_{c,t} \;\mapsto\; \tfrac{1}{2} \Big[
1 + \left( z_{a_c} \partial_{z_{a_c}} - z_{b_c} \partial_{z_{b_c}} \right) \\
+ \left( z_{a_t} \partial_{z_{a_t}} - z_{b_t} \partial_{z_{b_t}} \right)
- \left( z_{a_c} \partial_{z_{a_c}} - z_{b_c} \partial_{z_{b_c}} \right) \\
\times \left( z_{a_t} \partial_{z_{a_t}} - z_{b_t} \partial_{z_{b_t}} \right)
\Big].
\end{multline}
\item \textbf{SWAP$_{j,k}$}:
The SWAP gate admits the decomposition:
\begin{equation}
\mathrm{SWAP}_{j,k} = \tfrac{1}{2} \left( 1 + X_j X_k + Y_j Y_k + Z_j Z_k \right).
\end{equation}
Substituting the differential forms yields:
\begin{multline}
\mathrm{SWAP}_{j,k} \;\mapsto\; \tfrac{1}{2} \Big[
1 \\
+ \left( z_{a_j} \partial_{z_{b_j}} + z_{b_j} \partial_{z_{a_j}} \right)
\left( z_{a_k} \partial_{z_{b_k}} + z_{b_k} \partial_{z_{a_k}} \right) \\
- \left( z_{a_j} \partial_{z_{b_j}} - z_{b_j} \partial_{z_{a_j}} \right)
\left( z_{a_k} \partial_{z_{b_k}} - z_{b_k} \partial_{z_{a_k}} \right) \\
+ \left( z_{a_j} \partial_{z_{a_j}} - z_{b_j} \partial_{z_{b_j}} \right)
\left( z_{a_k} \partial_{z_{a_k}} - z_{b_k} \partial_{z_{b_k}} \right)
\Big].
\end{multline}
\item \textbf{General Controlled-$U$}:
For any single-qubit unitary $U = u_{00} \mathbb{I} + u_{01} X + u_{10} Y + u_{11} Z$, the controlled-$U$ gate is:
\begin{equation}
\Lambda_c(U_t) = \tfrac{1 + Z_c}{2} + \tfrac{1 - Z_c}{2} \, U_t,
\end{equation}
with Bargmann representation obtained by substituting the differential forms of $Z_c$ and $U_t$.
\end{itemize}
All multi-qubit gates preserve the local homogeneity condition $z_{a_j} \partial_{z_{a_j}} + z_{b_j} \partial_{z_{b_j}} = 1$ for every qubit $j$, ensuring they map physical states to physical states.

\section{Phasor Representation of Quantum Gates}
\label{sec:phasor}
We now consider the restriction of the Bargmann representation to the unit circle, where each holomorphic variable is expressed in phasor form as
\begin{equation}
z_{a_j} = e^{i\phi_{a_j}}, \quad z_{b_j} = e^{i\phi_{b_j}},
\end{equation}
with $\phi_{a_j}, \phi_{b_j} \in [0, 2\pi)$. This corresponds to fixing the magnitude $|z| = 1$ while retaining the phase degree of freedom, which is sufficient to represent qubit states due to the homogeneity constraint $z_{a_j}\partial_{z_{a_j}} + z_{b_j}\partial_{z_{b_j}} = 1$.
Under this restriction, the physical state space becomes a torus $\mathbb{T}^{2N}$ parameterized by the phase angles $\{\phi_{a_j}, \phi_{b_j}\}_{j=1}^N$. The basis states correspond to:
\begin{align}
\ket{0}_j \equiv \ket{\uparrow}_j &\mapsto e^{i\phi_{a_j}}, \\
\ket{1}_j \equiv \ket{\downarrow}_j &\mapsto e^{i\phi_{b_j}}.
\end{align}
The differential operators transform according to the chain rule:
\begin{equation}
\frac{\partial}{\partial z_{a_j}} = \frac{\partial \phi_{a_j}}{\partial z_{a_j}} \frac{\partial}{\partial \phi_{a_j}} = -\frac{i}{z_{a_j}} \frac{\partial}{\partial \phi_{a_j}} = -i e^{-i\phi_{a_j}} \frac{\partial}{\partial \phi_{a_j}},
\end{equation}
and similarly for $z_{b_j}$. Substituting these into the gate representations yields:
\subsection{Single-Qubit Gates in Phasor Form}
\begin{itemize}[left=0pt]
\item \textbf{Pauli-$X$ (bit-flip)}:
\begin{equation}
X_j \;\mapsto\; -i\left( e^{i(\phi_{a_j} - \phi_{b_j})} \frac{\partial}{\partial \phi_{b_j}} + e^{i(\phi_{b_j} - \phi_{a_j})} \frac{\partial}{\partial \phi_{a_j}} \right).
\end{equation}
\item \textbf{Pauli-$Y$ (bit-phase flip)}:
\begin{equation}
Y_j \;\mapsto\; -\left( e^{i(\phi_{a_j} - \phi_{b_j})} \frac{\partial}{\partial \phi_{b_j}} - e^{i(\phi_{b_j} - \phi_{a_j})} \frac{\partial}{\partial \phi_{a_j}} \right).
\end{equation}
\item \textbf{Pauli-$Z$ (phase flip)}:
\begin{equation}
Z_j \;\mapsto\; -i\left( \frac{\partial}{\partial \phi_{a_j}} - \frac{\partial}{\partial \phi_{b_j}} \right).
\end{equation}
\item \textbf{Hadamard gate}:
The Hadamard transformation becomes a rotation in Segal--Bargmann space:
\begin{multline}
(H_j f)(\dots, \phi_{a_j}, \phi_{b_j}, \dots) = \\
f\left(\dots, \arg\left(e^{i\phi_{a_j}} + e^{i\phi_{b_j}}\right), \arg\left(e^{i\phi_{a_j}} - e^{i\phi_{b_j}}\right), \dots \right),
\end{multline}
where $\arg(z)$ denotes the principal argument of $z$. Equivalently, defining $\Delta\phi_j = \phi_{a_j} - \phi_{b_j}$, we have:
\begin{align}
\phi_{a_j}' &= \arg\left(1 + e^{i\Delta\phi_j}\right) = \frac{\phi_{a_j} + \phi_{b_j}}{2} + \frac{\pi}{2} - \arctan\left(\cot\frac{\Delta\phi_j}{2}\right), \\
\phi_{b_j}' &= \arg\left(1 - e^{i\Delta\phi_j}\right) = \frac{\phi_{a_j} + \phi_{b_j}}{2} - \arctan\left(\tan\frac{\Delta\phi_j}{2}\right).
\end{align}
\item \textbf{Two-qubit SWAP gate}:
The SWAP operation exchanges phase angles between qubits:
\begin{multline}
(\mathrm{SWAP}_{jk} f)(\dots, \phi_{a_j}, \phi_{b_j}, \dots, \phi_{a_k}, \phi_{b_k}, \dots) \\
= f(\dots, \phi_{a_k}, \phi_{b_k}, \dots, \phi_{a_j}, \phi_{b_j}, \dots).
\end{multline}
\end{itemize}
\subsection{Geometric Interpretation on Toroidal Phase Space}
\label{subsec:geometric}
The restriction to unit-magnitude holomorphic variables $z_{a_j} = e^{i\phi_{a_j}}$, $z_{b_j} = e^{i\phi_{b_j}}$ maps the physical subspace of Segal-Bargmann space $\mathcal{H}_{\text{SB}}$ onto the $2N$-torus $\mathbb{T}^{2N} = (S^1)^{2N}$, where each $S^1$ corresponds to a phase angle. Quantum states are then represented as functions $f(\boldsymbol{\phi})$ on this torus, and quantum gates act as canonical transformations that preserve the symplectic structure induced by the homogeneity constraint.
\subsubsection{Hamiltonian Flow Definition}
To establish the geometric framework rigorously, we first define the Hamiltonian flow structure on $\mathbb{T}^{2N}$.
\begin{definition}[Hamiltonian Flow on $\mathbb{T}^{2N}$]
\label{def:hamiltonian_flow}
Let $(\mathbb{T}^{2N}, \omega)$ be the $2N$-dimensional torus equipped with the canonical symplectic form
\begin{equation}
\omega = \sum_{j=1}^N d\phi_{a_j} \wedge d\phi_{b_j}.
\label{eq:symplectic_form}
\end{equation}
For any smooth real-valued function $H: \mathbb{T}^{2N} \to \mathbb{R}$ (the Hamiltonian), the associated \textit{Hamiltonian vector field} $X_H$ is uniquely defined by the relation
\begin{equation}
\iota_{X_H} \omega = dH,
\label{eq:hamiltonian_vector_field}
\end{equation}
where $\iota_{X_H}$ denotes the interior product. In local coordinates $(\boldsymbol{\phi}_a, \boldsymbol{\phi}_b)$, this yields Hamilton's equations:
\begin{equation}
\dot{\phi}_{a_j} = \frac{\partial H}{\partial \phi_{b_j}}, \quad
\dot{\phi}_{b_j} = -\frac{\partial H}{\partial \phi_{a_j}}.
\label{eq:hamiltons_equations}
\end{equation}
The \textit{Hamiltonian flow} $\Phi_H^t: \mathbb{T}^{2N} \to \mathbb{T}^{2N}$ is the one-parameter family of diffeomorphisms generated by integrating $X_H$ over time $t$:
\begin{equation}
\frac{d}{dt} \Phi_H^t(\boldsymbol{\phi}) = X_H\big(\Phi_H^t(\boldsymbol{\phi})\big), \quad \Phi_H^0 = \text{id}.
\label{eq:flow_definition}
\end{equation}
\end{definition}
\begin{proposition}[Symplectic Properties]
\label{prop:symplectic}
The Hamiltonian flow $\Phi_H^t$ satisfies \cite{foundations, arnold}:
\begin{enumerate}
\item \textbf{Symplectomorphism:} $(\Phi_H^t)^* \omega = \omega$ (preserves symplectic structure)
\item \textbf{Energy Conservation:} $H \circ \Phi_H^t = H$ (Hamiltonian is constant along trajectories)
\item \textbf{Liouville's Theorem:} $\text{vol}(\Phi_H^t(U)) = \text{vol}(U)$ for any measurable set $U \subset \mathbb{T}^{2N}$
\item \textbf{Group Property:} $\Phi_H^{t+s} = \Phi_H^t \circ \Phi_H^s$
\end{enumerate}
\end{proposition}
These properties ensure that quantum gates represented as Hamiltonian flows are automatically unitary, as symplectic transformations on phase space correspond to unitary transformations in the quantum Hilbert space via the metaplectic representation.
\subsubsection{Pauli Gates as Hamiltonian Flows}
\begin{figure*}[t]
\centering
\begin{minipage}[b]{0.32\columnwidth}
\centering
\begin{tikzpicture}[scale=0.65]
\def\xmin{-1.8} \def\xmax{1.8}
\def\ymin{-1.8} \def\ymax{1.8}
\draw[->] (\xmin,0) -- (\xmax,0) node[right] {\scriptsize $\phi_a$};
\draw[->] (0,\ymin) -- (0,\ymax) node[above] {\scriptsize $\phi_b$};
\foreach \c in {-1.2,-0.6,0,0.6,1.2} {
\draw[blue!80!black, thick, ->]
({\xmin}, {\c - \xmin}) --
({\xmax}, {\c - \xmax});
}
\node[anchor=north] at (0,\ymin-0.2) {\scriptsize (a) $Z_j$};
\end{tikzpicture}
\end{minipage}
\hfill
\begin{minipage}[b]{0.32\columnwidth}
\centering
\begin{tikzpicture}[scale=0.65]
\def\xmin{-1.8} \def\xmax{1.8}
\def\ymin{-1.8} \def\ymax{1.8}
\draw[->] (\xmin,0) -- (\xmax,0) node[right] {\scriptsize $\phi_a$};
\draw[->] (0,\ymin) -- (0,\ymax) node[above] {\scriptsize $\phi_b$};
\foreach \c in {-1.2,-0.6,0,0.6,1.2} {
\draw[red!80!black, thick, domain=-1.7:1.7, smooth, samples=40, ->]
plot ({\x}, {\c - \x + 0.3*sin(2*deg(\x))});
}
\fill[red!80!black] (0.785,0) circle (1.2pt);
\fill[red!80!black] (-0.785,0) circle (1.2pt);
\node[anchor=north] at (0,\ymin-0.2) {\scriptsize (b) $X_j$};
\end{tikzpicture}
\end{minipage}
\hfill
\begin{minipage}[b]{0.32\columnwidth}
\centering
\begin{tikzpicture}[scale=0.65]
\def\xmin{-1.8} \def\xmax{1.8}
\def\ymin{-1.8} \def\ymax{1.8}
\draw[->] (\xmin,0) -- (\xmax,0) node[right] {\scriptsize $\phi_a$};
\draw[->] (0,\ymin) -- (0,\ymax) node[above] {\scriptsize $\phi_b$};
\foreach \c in {-1.2,-0.6,0,0.6,1.2} {
\draw[green!70!black, thick, domain=-1.7:1.7, smooth, samples=40, ->]
plot ({\x}, {\c - \x + 0.3*cos(2*deg(\x))});
}
\fill[green!70!black] (0,0) circle (1.2pt) node[below right] {\tiny $\Delta\phi=0$};
\fill[green!70!black] (1.57,0) circle (1.2pt) node[below right] {\tiny $\Delta\phi=\pi$};
\node[anchor=north] at (0,\ymin-0.2) {\scriptsize (c) $Y_j$};
\end{tikzpicture}
\end{minipage}
\caption{
Phase portraits of Pauli gate dynamics on the $(\phi_{a_j}, \phi_{b_j})$ plane.
(a) $Z_j$: Uniform translation along $\Delta\phi_j = \phi_{a_j} - \phi_{b_j}$ (straight lines of constant $\Sigma\phi_j$).
(b) $X_j$: Nonlinear oscillations with fixed points at $\Delta\phi_j = \pm\pi/2$ (eigenstates $\ket{\pm}$).
(c) $Y_j$: Oscillations with fixed points at $\Delta\phi_j = 0$ and $\Delta\phi_j = \pi$ (eigenstates $\ket{+i}$ and $\ket{-i}$, respectively).
All flows preserve $\Sigma\phi_j = \phi_{a_j} + \phi_{b_j}$, reflecting the homogeneity constraint.
Trajectories follow Hamilton's equations~(\ref{eq:hamiltons_equations}) with Hamiltonians specified in Definition~\ref{def:hamiltonian_flow}.
}
\label{fig:pauli_flows}
\end{figure*}
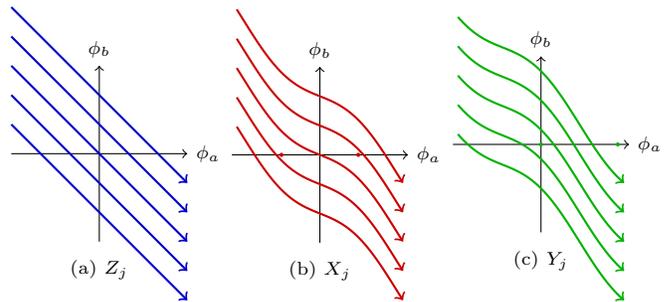

Each Pauli operator generates a Hamiltonian flow on the toroidal phase space $\mathbb{T}^{2N}$, governed by the symplectic form in Eq.~(\ref{eq:symplectic_form}). The Pauli operators correspond to specific Hamiltonian functions:
\begin{itemize}[left=0pt]
\item \textbf{$Z_j$ gate (Phase rotation):} The Hamiltonian is $H_{Z_j} = \phi_{a_j} - \phi_{b_j} = \Delta\phi_j$. From Eq.~(\ref{eq:hamiltons_equations}), the equations of motion are:
\begin{equation}
\dot{\phi}_{a_j} = \frac{\partial H_{Z_j}}{\partial \phi_{b_j}} = -1, \quad
\dot{\phi}_{b_j} = -\frac{\partial H_{Z_j}}{\partial \phi_{a_j}} = 1.
\label{eq:Z_equations}
\end{equation}
Integrating yields straight-line trajectories:
\begin{equation}
\phi_{a_j}(t) = \phi_{a_j}(0) - t, \quad
\phi_{b_j}(t) = \phi_{b_j}(0) + t.
\label{eq:Z_solution}
\end{equation}
Thus, $Z_j$ generates uniform translation along the relative phase $\Delta\phi_j$, while the total phase $\Sigma\phi_j = \phi_{a_j} + \phi_{b_j}$ remains constant, corresponding to pure phase accumulation without population transfer. The flow period is $T_Z = 2\pi$, matching the $2\pi$ periodicity of the torus.
\item \textbf{$X_j$ gate (Bit flip):} The Hamiltonian is $H_{X_j} = \sin(\Delta\phi_j)$. The equations of motion are:
\begin{equation}
\dot{\phi}_{a_j} = \cos(\Delta\phi_j), \quad
\dot{\phi}_{b_j} = -\cos(\Delta\phi_j),
\label{eq:X_equations}
\end{equation}
yielding coupled oscillations in the $(\phi_{a_j}, \phi_{b_j})$ plane. In sum and difference coordinates:
\begin{equation}
\dot{\Sigma\phi}_j = 0, \quad
\dot{\Delta\phi}_j = 2\cos(\Delta\phi_j).
\label{eq:X_coordinates}
\end{equation}
This describes periodic exchange of amplitude between the two modes. The flow lines are level sets of $\Sigma\phi_j$, with fixed points at $\Delta\phi_j = \pm \pi/2$ (corresponding to eigenstates $\ket{\pm}$ of $X_j$). The period depends on initial conditions:
\begin{equation}
T_X(\Delta\phi_0) = \int_0^{2\pi} \frac{d\theta}{2\cos\theta} = \pi \sec(\Delta\phi_0).
\label{eq:X_period}
\end{equation}
\item \textbf{$Y_j$ gate:} Similarly, $Y_j$ corresponds to the Hamiltonian $H_{Y_j} = -\cos(\Delta\phi_j)$, generating flows with equations:
\begin{equation}
\dot{\Delta\phi}_j = 2\sin(\Delta\phi_j),
\label{eq:Y_equations}
\end{equation}
with fixed points at $\Delta\phi_j = 0$ and $\Delta\phi_j = \pi$ (modulo $2\pi$). These correspond to the eigenstates of $Y_j$: $\ket{+i} = (\ket{0} + i\ket{1})/\sqrt{2}$ at $\Delta\phi_j = 0$ and $\ket{-i} = (\ket{0} - i\ket{1})/\sqrt{2}$ at $\Delta\phi_j = \pi$. The flow trajectories connect these fixed points along meridians of the torus, as illustrated in Fig.~\ref{fig:pauli_flows}(c).
\end{itemize}
These flows are geodesics of the flat metric $g = \sum_j (d\phi_{a_j}^2 + d\phi_{b_j}^2)$ on $\mathbb{T}^{2N}$, reflecting the Abelian nature of the underlying bosonic algebra. The distinct dynamical signatures of each Pauli gate are illustrated in Fig.~\ref{fig:pauli_flows}. Moreover, the Poisson bracket
\begin{equation}
\{H_A, H_B\} = \omega(X_A, X_B) = \sum_j \left( \frac{\partial H_A}{\partial \phi_{a_j}} \frac{\partial H_B}{\partial \phi_{b_j}} - \frac{\partial H_A}{\partial \phi_{b_j}} \frac{\partial H_B}{\partial \phi_{a_j}} \right)
\label{eq:poisson_bracket}
\end{equation}
reproduces the $\mathfrak{su}(2)$ Lie algebra: $\{H_X, H_Y\} = H_Z$, $\{H_Y, H_Z\} = H_X$, $\{H_Z, H_X\} = H_Y$, confirming that the Pauli gates realize the spin algebra as canonical transformations on phase space.
\subsubsection{Hadamard as a Nonlinear Automorphism}
The Hadamard transformation induces a diffeomorphism $\mathcal{H}_j : \mathbb{T}^2 \to \mathbb{T}^2$ on the 2-torus associated with qubit $j$:
\begin{equation}
\begin{pmatrix} \phi_{a_j}' \\ \phi_{b_j}' \end{pmatrix}
=
\begin{pmatrix}
\arg\!\big(1 + e^{i\Delta\phi_j}\big) \\
\arg\!\big(1 - e^{i\Delta\phi_j}\big)
\end{pmatrix}
+ \frac{\Sigma\phi_j}{2} \begin{pmatrix} 1 \\ 1 \end{pmatrix},
\label{eq:hadamard_map}
\end{equation}
where $\Sigma\phi_j = \phi_{a_j} + \phi_{b_j}$ and $\Delta\phi_j = \phi_{a_j} - \phi_{b_j}$. This map is:
\begin{itemize}
\item \textit{Nonlinear:} The output phases depend nontrivially on $\Delta\phi_j$ via trigonometric functions.
\item \textit{Singular:} At $\Delta\phi_j = \pi \mod 2\pi$, the argument becomes undefined (corresponding to the state $\ket{1}$ mapping to $\ket{-}$, which has a phase ambiguity).
 \item \textit{Area-preserving:} The Jacobian matrix $D\mathcal{H}_j$ of the transformation has determinant $\det(D\mathcal{H}_j) = 1$. 
To demonstrate this, we express $\mathcal{H}_j$ in the sum-and-difference coordinates $(\Sigma\phi_j, \Delta\phi_j)$, where the transformation acts as $(\Sigma\phi_j, \Delta\phi_j) \mapsto (\Sigma\phi_j, \Delta\phi_j')$ with
\begin{equation}
\Delta\phi_j' = \arg\!\big(1+e^{i\Delta\phi_j}\big) - \arg\!\big(1-e^{i\Delta\phi_j}\big).
\end{equation}
Since $\Sigma\phi_j$ is invariant, the Jacobian determinant reduces to the scalar derivative $J = \partial\Delta\phi_j'/\partial\Delta\phi_j$. 
Using the identity $\frac{d}{d\theta}\arg f(\theta) = \operatorname{Im}\!\big[f'(\theta)/f(\theta)\big]$ for a complex-valued function $f$, we compute:
\begin{align}
\frac{d}{d\theta}\arg\!\big(1+e^{i\theta}\big) &= \operatorname{Im}\!\left[\frac{ie^{i\theta}}{1+e^{i\theta}}\right] = \frac{1}{2}, \\
\frac{d}{d\theta}\arg\!\big(1-e^{i\theta}\big) &= \operatorname{Im}\!\left[\frac{-ie^{i\theta}}{1-e^{i\theta}}\right] = -\frac{1}{2},
\end{align}
where the second result follows from simplifying the imaginary part of the quotient. 
Consequently, the total derivative is
\begin{equation}
J = \frac{d\Delta\phi_j'}{d\Delta\phi_j} = \frac{1}{2} - \left(-\frac{1}{2}\right) = 1.
\end{equation}
This confirms that the Hadamard map preserves the symplectic volume form $d\Sigma\phi_j \wedge d\Delta\phi_j$ on the toroidal phase space.
\end{itemize}
Thus, Hadamard implements a nonlinear shear on the torus, mixing the sum and difference coordinates in a phase-dependent way. Unlike Pauli gates, Hadamard cannot be generated by a time-independent Hamiltonian flow; it requires a time-dependent protocol or instantaneous transformation.
\subsubsection{Multi-qubit Gates as Entangling Diffeomorphisms}
Gates like CNOT and SWAP induce correlations between distinct toroidal factors. For example:
\begin{itemize}
\item \textbf{SWAP} acts as a permutation of toroidal coordinates: $(\phi_{a_j}, \phi_{b_j}, \phi_{a_k}, \phi_{b_k}) \mapsto (\phi_{a_k}, \phi_{b_k}, \phi_{a_j}, \phi_{b_j})$, which is a global isometry of $\mathbb{T}^{2N}$.
\item \textbf{CNOT} implements a conditional flow: the vector field on the target qubit's torus is activated only when the control qubit's phase satisfies $\phi_{b_c} = 0 \mod 2\pi$ (i.e., the control is in $\ket{0}$). This creates a fiber bundle structure over the control torus, with the target dynamics twisting along the base. The Hamiltonian can be written as:
\begin{equation}
H_{\text{CNOT}} = \Theta(\phi_{b_c}) \cdot H_{X_t},
\label{eq:cnot_hamiltonian}
\end{equation}
where $\Theta$ is the Heaviside step function and $H_{X_t}$ acts on the target qubit.
\end{itemize}
These operations generate the full group of volume-preserving diffeomorphisms on $\mathbb{T}^{2N}$, enabling universal quantum computation through geometric manipulation of phase space.
\subsubsection{Topological Considerations}
This geometric perspective reveals that quantum computation can be viewed as the controlled navigation of a point (or wavefunction) on a high-dimensional torus, where gates correspond to carefully engineered flows and deformations. The topological properties of $\mathbb{T}^{2N}$, such as its nontrivial fundamental group $\pi_1(\mathbb{T}^{2N}) = \mathbb{Z}^{2N}$, underlie the robustness of quantum phases and enable topological error correction schemes~\cite{Hatcher,Pachos}.
The winding numbers $(n_{a_1}, n_{b_1}, \dots, n_{a_N}, n_{b_N}) \in \mathbb{Z}^{2N}$ classify distinct homotopy classes of trajectories, providing topological invariants that can be used for error detection. A computational error corresponds to an unintended change in winding numbers, which can be detected through interference measurements.
Furthermore, the restriction to $|z| = 1$ isolates the essential phase degrees of freedom, providing a minimal yet complete description of qubit dynamics in terms of classical-like phase space trajectories. This representation is particularly useful for analyzing phase coherence properties of quantum states, topological features of quantum circuits, semiclassical limits where phase dynamics dominate, and geometric phases (Berry phases) accumulated during computation. The unit magnitude constraint $|z| = 1$ effectively projects the full Segal-Bargmann space onto its boundary, capturing the essential phase information while eliminating redundant amplitude degrees of freedom. This provides a minimal representation sufficient for describing all quantum computational processes within the physical subspace.

\section{Beyond the Torus: Geometric Structures in Holomorphic Quantum Computation}
\label{sec:beyond_torus}
While the toroidal restriction $|z_{a_j}| = |z_{b_j}| = 1$ (i.e., $z = e^{i\phi}$) captures the essential phase dynamics of unitary quantum gates, it discards amplitude information critical for non-unitary processes (e.g., decoherence, measurement, dissipation), state preparation, and gradient-based optimization in variational quantum algorithms. The full Segal-Bargmann space $\mathcal{H}_{\text{SB}}$ carries a natural Kähler structure that restores this missing information and reveals deeper geometric features of quantum computation.
\subsection{Amplitude Dynamics in Full Kähler Space}
The Segal-Bargmann space $\mathcal{H}_{\text{SB}}$ is endowed with a Kähler potential
\begin{equation}
K(\mathbf{z}, \bar{\mathbf{z}}) = \|\mathbf{z}\|^2 = \sum_{j=1}^N \left( |z_{a_j}|^2 + |z_{b_j}|^2 \right),
\end{equation}
which induces a compatible triple of geometric structures:
\begin{itemize}[left=0pt]
\item A \textbf{Riemannian metric} $g_{i\bar{j}} = \partial_i \partial_{\bar{j}} K = \delta_{ij}$, which defines distances between quantum states via the fidelity $F(\psi, \phi) = |\langle \psi | \phi \rangle|^2$.
\item A \textbf{symplectic form} $\omega = i \sum_j dz_j \wedge d\bar{z}_j$, which governs Hamiltonian evolution through the equation $\dot{z}_j = -i \partial_{\bar{z}_j} H$.
\item A \textbf{complex structure} $J$ (multiplication by $i$), which encodes the uncertainty principle.
\end{itemize}
In non-unitary scenarios (e.g., Lindblad dynamics), the amplitude degrees of freedom $|z_j|$ evolve according to dissipative flows on $\mathbb{C}^{2N}$, while the toroidal subspace $\mathbb{T}^{2N}$ remains invariant only under purely unitary evolution. Thus, simulations of open quantum systems require the full Kähler geometry to capture relaxation toward fixed points (e.g., thermal states).
\subsection{Entanglement Geometry via Segre Embedding}
\label{subsec:entanglement_geometry}

The physical state space for $N$ qubits is the complex projective space $\mathbb{C}P^{2^N - 1}$, equipped with the canonical Fubini--Study metric \cite{Provost,geometry}
\begin{equation}
ds^2_{\mathrm{FS}} = \frac{\langle d\psi | d\psi \rangle}{\langle \psi | \psi \rangle} - \frac{|\langle \psi | d\psi \rangle|^2}{\langle \psi | \psi \rangle^2},
\end{equation}
which induces the quantum fidelity distance $D_{\mathrm{FS}}([\psi], [\phi]) = \arccos \sqrt{|\langle \psi | \phi \rangle|^2 / (\langle \psi | \psi \rangle \langle \phi | \phi \rangle)}$. Within this manifold, separable (unentangled) states form a lower-dimensional algebraic subvariety known as the \textit{Segre variety} \cite{Heydari,Holweck}:
\begin{equation}
\Sigma_N = \sigma\big(\underbrace{\mathbb{C}P^1 \times \cdots \times \mathbb{C}P^1}_{N \text{ times}}\big) \hookrightarrow \mathbb{C}P^{2^N - 1},
\end{equation}
where $\sigma$ denotes the Segre embedding $\sigma([v_1], \dots, [v_N]) = [v_1 \otimes \cdots \otimes v_N]$. 

In the Segal--Bargmann holomorphic representation, a state $f(\mathbf{z})$ is separable if and only if it factorizes into linear forms across qubit subsystems:
\begin{equation}
f(\mathbf{z}) = \prod_{j=1}^N \left( \alpha_j z_{a_j} + \beta_j z_{b_j} \right).
\end{equation}
This factorization criterion is the analytic manifestation of the algebraic condition that separable states correspond to rank-1 tensors in the underlying Hilbert space. Entangling unitaries act as diffeomorphisms that transport points off $\Sigma_N$ into the ambient projective space. For instance, the CNOT gate induces the map
\begin{equation}
\mathrm{CNOT}_{c,t} : \Sigma_N \to \mathbb{C}P^{2^N - 1} \setminus \Sigma_N,
\end{equation}
transforming the separable holomorphic function $f = z_{a_c} z_{a_t}$ into
\begin{equation}
f' = z_{a_c} z_{a_t} + z_{b_c} z_{b_t},
\label{eq:bell_holomorphic}
\end{equation}
which represents the Bell state $\ket{\Phi^+} = (\ket{00} + \ket{11})/\sqrt{2}$ up to projective normalization.  Explicitly, $z_{a_c}z_{a_t} \mapsto \ket{00}$ and $z_{b_c}z_{b_t} \mapsto \ket{11}$, so $f' \propto \ket{00} + \ket{11} = \sqrt{2}\ket{\Phi^+}$.
The degree of entanglement admits a natural geometric quantification via the minimal Fubini--Study distance to the Segre variety:
\begin{equation}
\mathcal{E}_{\mathrm{FS}}([\psi]) = \min_{[\phi] \in \Sigma_N} D_{\mathrm{FS}}([\psi], [\phi]).
\end{equation}
This geometric measure is directly related to the standard geometric entanglement monotone $E_G([\psi]) = -\log_2 \max_{[\phi] \in \Sigma_N} |\langle \psi | \phi \rangle|^2$ via the identity $\mathcal{E}_{\mathrm{FS}}([\psi]) = \arccos \sqrt{2^{-E_G([\psi])}}$. Crucially, $\mathcal{E}_{\mathrm{FS}}([\psi]) = 0$ if and only if $[\psi] \in \Sigma_N$, providing a rigorous metric characterization of separability.

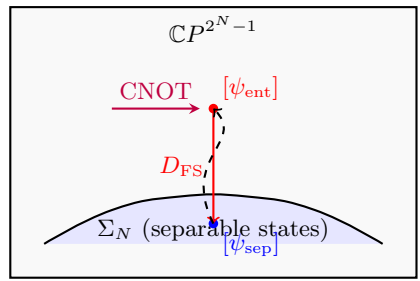
\begin{figure}[t]
\centering
\begin{tikzpicture}[scale=0.9]
\draw[fill=gray!5, thick] (-3,-2) rectangle (3,2);
\node[anchor=north] at (0,2) {\small $\mathbb{C}P^{2^N - 1}$};
\draw[fill=blue!10, thick, smooth] plot [tension=0.8] coordinates {
(-2.5,-1.5) (-1.5,-1) (-0.5,-0.8) (0.5,-0.8) (1.5,-1) (2.5,-1.5)
};
\node[anchor=south] at (0,-1.6) {\small $\Sigma_N$ (separable states)};
\fill[blue] (0,-1.2) circle (2pt) node[below right] {\small $[\psi_{\text{sep}}]$};
\fill[red] (0,0.5) circle (2pt) node[above right] {\small $[\psi_{\text{ent}}]$};
\draw[red, thick, ->] (0,0.5) -- (0,-1.2)
node[midway, left] {\small $D_{\mathrm{FS}}$};
\draw[->, >=stealth, thick, purple] (-1.5,0.5) -- (-0.2,0.5)
node[midway, above] {\small CNOT};
\draw[->, dashed, thick] (0,-1.2) .. controls (-0.5,-0.2) and (0.5,0) .. (0,0.5);
\node[anchor=north, text width=7cm, align=center] at (0,-2.2) {
};
\end{tikzpicture}
\caption{
Schematic illustration of entanglement geometry in projective Hilbert space.
The Segre variety $\Sigma_N$ (blue submanifold) contains all separable states.
An entangling gate maps a separable state $[\psi_{\text{sep}}]$ to an entangled state
$[\psi_{\text{ent}}]$, with the Fubini--Study distance $D_{\mathrm{FS}}$ measuring
the minimal geodesic distance to $\Sigma_N$. The Fubini--Study distance
$D_{\mathrm{FS}}$ quantifies the degree of entanglement.
}
\label{fig:fubini_study}
\end{figure}
\subsection{Topological Protection from Fiber Bundle Structure}
The constraint $|z_{a_j}|^2 + |z_{b_j}|^2 = 1$ for each qubit defines a Hopf fibration \cite{Hatcher}:
\begin{equation}
S^1 \hookrightarrow S^3 \xrightarrow{\pi} S^2,
\end{equation}
where the total space $S^3$ is the unit sphere in $\mathbb{C}^2$ (physical states before projectivization), the base space $S^2$ is the Bloch sphere (physical qubit states), and the fiber $S^1$ represents the unobservable global phase. For $N$ qubits, this generalizes to a principal $U(1)^N$-bundle:
\begin{equation}
U(1)^N \hookrightarrow (S^3)^N \xrightarrow{\pi} (S^2)^N.
\end{equation}
Quantum gates that act locally (e.g., single-qubit rotations) correspond to vertical automorphisms (they move only along fibers), while entangling gates induce horizontal lifts that twist the bundle. Crucially, the holonomy of this bundle, computed via the Berry connection
\begin{equation}
\mathcal{A} = i \sum_j \left( \bar{z}_{a_j} dz_{a_j} + \bar{z}_{b_j} dz_{b_j} \right)
\end{equation}
is robust against local phase noise. This explains why topological quantum computations are inherently fault-tolerant: errors that act only on the fiber (phase fluctuations) do not affect the base-space computation (the logical qubit state).
\subsection{Semiclassical Limits and Coherent State Path Integrals}
The Kähler potential $K = \|\mathbf{z}\|^2$ enables a direct connection to Feynman path integrals via coherent states \cite{Klauder,Kirwin,Ali}. The transition amplitude between states $|\psi_i\rangle$ and $|\psi_f\rangle$ is given by:
\begin{multline}
\langle \psi_f | e^{-iHt} | \psi_i \rangle = \int \mathcal{D}[\bar{z}, z]  \\
\times \exp\left( i \int_0^t \left[ \frac{i}{2} (\bar{z} \dot{z} - \dot{\bar{z}} z) - H(\bar{z}, z) \right] dt' \right),
\end{multline}
where the measure $\mathcal{D}[\bar{z}, z]$ is induced by the Kähler metric. In the semiclassical limit ($\hbar \to 0$), the path integral is dominated by classical trajectories satisfying:
\begin{align}
i \dot{z}_j &= \frac{\partial H}{\partial \bar{z}_j}, \\
-i \dot{\bar{z}}_j &= \frac{\partial H}{\partial z_j}.
\end{align}
These are precisely the Hamiltonian equations on the Kähler manifold. For the gate representations developed in this work:
\begin{itemize}[left=0pt]
\item Pauli gates correspond to quadratic Hamiltonians, e.g., $H_X = \frac{1}{2}(z_a \bar{z}_b + z_b \bar{z}_a)$,
\item The Hadamard gate arises from a time-dependent linear Hamiltonian.
\end{itemize}
This framework allows semiclassical simulation of quantum circuits by solving Hamiltonian ordinary differential equations (ODEs) on $\mathbb{C}^{2N}$, with quantum corrections computable via loop expansions in powers of $\hbar$. Specifically, the path integral
\begin{equation}
\langle \psi_f | e^{-iHt} | \psi_i \rangle = \int \mathcal{D}[\bar{z}, z]  \exp\left( \frac{i}{\hbar} S[\bar{z}, z] \right)
\end{equation}
is evaluated by expanding the action $S[\bar{z}, z] = \int_0^t \left[ \frac{i\hbar}{2} (\bar{z} \dot{z} - \dot{\bar{z}} z) - H(\bar{z}, z) \right] dt'$ around its classical stationary point $(\bar{z}_{\mathrm{cl}}, z_{\mathrm{cl}})$ satisfying the Kähler Hamilton equations. The leading-order contribution yields the semiclassical propagator
\begin{equation}
K_{\mathrm{sc}}(t) = \mathcal{N} \exp\left( \frac{i}{\hbar} S_{\mathrm{cl}} \right),
\end{equation}
where $S_{\mathrm{cl}} = S[\bar{z}_{\mathrm{cl}}, z_{\mathrm{cl}}]$ and $\mathcal{N}$ is a prefactor determined by Gaussian fluctuations. Higher-order quantum corrections arise from evaluating functional determinants of the fluctuation operator, which correspond to loop diagrams in the associated field theory. For weakly entangled states, those remaining near the Segre variety $\Sigma_N$ during evolution, these fluctuations are suppressed, making the semiclassical approximation particularly accurate and computationally efficient compared to full Hilbert space simulations.
\section{Conclusion}\label{sec:conclusion}
We have established a comprehensive framework for representing quantum logic gates in the Segal--Bargmann space. First, we derived explicit differential operator forms for fundamental single- and multi-qubit gates, including Pauli operators, Hadamard, SWAP, CNOT, and CZ, acting on holomorphic variables $z_{a_j}, z_{b_j}$. Crucially, we demonstrated that physical qubit states correspond to functions homogeneous of degree one in each bosonic pair, and all quantum gates preserve this constraint while implementing unitary transformations. The restriction to unit-magnitude variables ($|z| = 1$) reveals a toroidal space $\mathbb{T}^{2N}$ where gates act as canonical transformations: Pauli operators generate Hamiltonian flows, Hadamard implements a nonlinear automorphism, and entangling gates create correlations between distinct toroidal factors. Beyond the torus, the full Kähler geometry of the Segal--Bargmann space provides a richer structure that captures amplitude dynamics essential for non-unitary processes, while the Segre embedding offers a geometric characterization of entanglement as deviation from separable manifolds. The underlying fiber bundle structure explains topological protection against phase noise, and the Kähler potential enables semiclassical simulations via coherent state path integrals.

\end{document}